# TIME-FREQUENCY ANALYSIS OF GALLEX AND GNO SOLAR NEUTRINO DATA: EVIDENCE SUGGESTIVE OF ASYMMETRIC AND VARIABLE NUCLEAR BURNING

P.A. STURROCK

*Center for Space Science and Astrophysics, Stanford University,*

*Stanford, California 94305-4060, U.S.A.*

*(email: sturrock@stanford.edu)*

**Abstract.** Time-frequency analysis of data from the GALLEX and GNO solar neutrino experiments shows that some features in power-spectrum analyses of those datasets are due to aliasing (a result of the fact that run durations tend to be small multiples of one week). Displays formed from the published GALLEX data show a sharp discontinuity that we attribute to some systematic effect. We therefore normalize data for each of the four experiments in the GALLEX series and concatenate the resulting normalized data. This step effectively removes the presumably systematic effect. To help understand the effect of aliasing, we form time-frequency displays of the two principal modulations found in the data, at 11.87 yr$^{-1}$ and at 13.63 yr$^{-1}$. We also form time-frequency displays of datasets formed by subtracting these modulations from the actual (normalized) data. The results suggest that the true modulation is that at 11.87 yr$^{-1}$. This result is consistent with the results of our earlier analysis of Homestake solar neutrino data, which points to modulations originating in a region where the sidereal rotation rate is close to 13.87 yr$^{-1}$. Comparison with helioseismology data indicates that modulation is occurring either in the radiative zone just below the tachocline, presumably by the RSFP (resonant spin-flavor precession) process, or in the core, presumably due to inhomogeneities and fluctuations in the nuclear-burning process.

# 1. Introduction

In previous articles, we have carried out power-spectrum analyses of Homestake (Sturrock, Walther, and Wheatland 1997), GALLIUM-GNO (Sturrock, Scargle, Walther, and Wheatland, 1999; Sturrock and Scargle, 2001; Sturrock and Weber, 2002; Sturrock, Caldwell, and Scargle, 2006), Super-Kamiokande (Sturrock, 2003, 2004; Sturrock, Caldwell, Scargle, and Wheatland, 2005; Sturrock and Scargle, 2006), and SNO (Sturrock, 2006) datasets. In this article, we re-analyze GALLEX (Anselmann 1993, 1995; Hampel 1996, 1999) and GNO (Altmann 2000, 2005; Kirsten 2003) data. We plan to re-analyze the other datasets in later articles.

We have recently re-visited the question of whether capture-rate measurements made by the GALLEX and GNO experiments are unimodal or bimodal, and find strong evidence for bimodality of the GALLEX measurements (Sturrock, 2008a). This is strong evidence for variability, but the variability could conceivably be due to the experimental procedures rather than to the neutrino flux itself. If we adopt the results at face value, we face the difficulty that the upper lobe in the distribution, corresponding to a capture rate of $111.2 \pm 13.9\, SNU$, is incompatible with the LMA version of the MSW process, on the assumption that the neutrino-production process is steady (Pulido 2007). However, the difficulty disappears if we allow for the possibility that the neutrino-production process may not be steady.

We have found evidence for variability of the GALLEX measurements on a time-scale of a year or so (Sturrock, Caldwell, and Scargle, 2006), but this may be due in part to systematic changes between the four separate GALLEX experiments. This article also presents the results of power-spectrum analysis of the datasets: for the frequency range $0-26\, yr^{-1}$, the strongest peak is found at $13.64\, yr^{-1}$. It was tempting to attribute this feature to modulation of the neutrino flux by a magnetic structure located where the sidereal rotation rate is $14.64\, yr^{-1}$, i.e. at about 0.75 $R_{solar}$ (Schou et al. 1998). This falls low in the convection zone, where the magnetic-field strength may be several hundred thousand gauss. This scenario requires that neutrinos possess a magnetic moment, or a transition magnetic moment (Akhmedov 1988; Lim and Marciano, 1988; Chauhan,

Pulido, and Raghavan 2005). This scenario led to the conjecture that – in addition to the known neutrino flavors – there is an additional sterile neutrino (Caldwell 2007).

However, power-spectrum analysis of radio-chemical solar-neutrino data is complicated – indeed compromised – by the fact that runs are typically scheduled to have durations that are small multiples (typically 3 or 4) of one week. This unfortunately leads to severe aliasing in any power-spectrum analysis. Furthermore, due to variability associated with the small number of counts per run, the strength of a peak in a power spectrum is quite variable, even for a fixed value of the amplitude. For these reasons, it is helpful to look for another method of analysis that is more informative, and may therefore help determine whether a given peak is intrinsic or merely an alias of another peak. It is for this reason that we have carried out time-frequency analyses of the GALLEX and GNO datasets.

In Section 2, we show the results of a time-frequency analysis of a dataset formed by merging the published results of all four GALLEX experiments. We find sharp discontinuities in the GALLEX display, indicating that there are systematic differences between the four experiments. It is therefore necessary to devise a scheme for combining data from the four experiments that allows for such systematic differences. In Section 3, we introduce a normalization process that is based on our recent procedure for an extended chi-square statistic (Sturrock 2008b). For each experiment, we convert the capture-rate estimate for each run to a proxy variable that has the property that if the neutrino flux were constant, with a value obtained from the maximum-likelihood procedure, and if the error were accurately represented by the quoted upper and lower error estimates, then the proxy would conform to a normal distribution with mean zero and standard deviation unity.

In Section 4, we carry out power-spectrum analyses of the GALLEX and GNO datasets, using the normalized measurements. The strongest peak in the range $0-30$ $yr^{-1}$ is that at 13.63 $yr^{-1}$, essentially unchanged. The second strongest peak is found at 11.87 $yr^{-1}$, and this raises the question of whether one peak is an alias of the other, since the sum of the two frequencies is 25.50 $yr^{-1}$, corresponding to a period of 14.3 days – close to two weeks. In Section 5, we carry out a power-spectrum analysis of the mean-times of runs and list the frequencies of the salient peaks. We find that there are indeed strong peaks near 26.1 $yr^{-1}$, the frequency corresponding to a period of two weeks, strengthening our suspicion that the peaks at 11.87 $yr^{-1}$ and at 13.63 $yr^{-1}$ are related. This poses the important question of which represents true modulation and which is merely an alias. If we were to assume that the modulation must occur in the convection zone, we would conclude that the true modulation is that at 13.63 $yr^{-1}$. However, it seems prudent not to jump to that conclusion.

In Section 6, we carry out a new time-frequency analysis of the GALLEX and GNO datasets, now using the normalized measurements. We find that this transformation has been helpful in that the sharp discontinuity in the display that we found in Section 2 does not now occur. Inspection of this display makes it quite clear that the peaks at 11.87 $yr^{-1}$ and at 13.63 $yr^{-1}$ are indeed an alias pair. It also suggests that some of the other features in the power spectrum are due to aliasing.

In Section 7, we show the results of time-frequency analyses of time series formed from hypothetical sinusoidal modulation of the solar neutrino flux, adopting the frequency and phase corresponding to the peaks at 11.87 $yr^{-1}$ and at 13.63 $yr^{-1}$. We also show the results of time-frequency analyses of time series formed by subtracting from the real data modulations corresponding to the peaks at 11.87 $yr^{-1}$ and at 13.63 $yr^{-1}$. We also carry out corresponding analyses of the GNO dataset, focusing on the peak at 9.20 $yr^{-1}$, since this is close to the principal peak (at 9.43 $yr^{-1}$) in the Super-Kamiokande dataset.

The implications of these results are discussed in Section 8.

## 2. Time-Frequency Analysis

It is curious and notable that power spectra obtained from the GALLEX dataset and from the GNO dataset are quite different (Sturrock, Caldwell and Scargle 2006). For the frequency range 0 to 26 yr$^{-1}$, the former contains five peaks with power $S \geq 5$ (at 4.54, 6.93, 11.87, 13.08, and 13.64 yr$^{-1}$), whereas the latter has only one such peak (at 21.93 yr$^{-1}$). The most prominent peak in the GALLEX power spectrum is that at 13.64 yr$^{-1}$ (with power 7.81), which we have previously interpreted as modulation occurring in the convection zone. However, it is obviously important to examine how the power spectrum evolves in time. This leads us to analyze the data by a time-frequency procedure.

We introduce a "sample length" L. If the complete dataset has R runs, we form in turn power spectra for runs 1 through L, then 2 through L + 1, etc., ending with R - L + 1 through R. We then display the power by color code in a map with the center-time of each sequence as abscissa, and frequency as ordinate.

We have first applied this procedure to the GALLEX and GNO datasets, with the choice L = 20, using the likelihood procedure for power spectrum analysis used in Sturrock, Caldwell and Scargle (2006). We show results for GALLEX in Figure 1. The sharp change at central time 1994 clearly indicates that there is an artifact in the data, which it is necessary to identify and correct for. We therefore defer analysis of GNO data.

### 3. Normalized Data

The GALLEX data were acquired in four separate experiments. The means and standard deviations of the count rate for each experiment are shown in Table 1. We see that there are sharp changes from one experiment to the next. For power spectrum analysis, it is particularly significant that the variance changes markedly from experiment to experiment, since this has a big effect on the power spectrum.

It is clearly necessary to normalize the data in such a way as to obviate these presumably systematic effects. It is convenient to use the same procedure that we used to extend the chi-square procedure for use with data that have non-normal distributions (Sturrock 2008b).

Given the best estimate $g_r$ of the count rate and the upper and lower error estimates $\sigma_{u,r}$ and $\sigma_{l,r}$, respectively, this information concerning the actual count rate $f$ is represented by the probability distribution function (pdf)

$$p_r(f)df = \frac{2^{1/2}}{\pi^{1/2}(\sigma_{u,r}+\sigma_{l,r})} \exp\left(-\frac{(f-g_r)^2}{2\sigma_{u,r}^2}\right) df \text{ for } f > g_r,$$

$$p_r(f)df = \frac{2^{1/2}}{\pi^{1/2}(\sigma_{u,r}+\sigma_{l,r})} \exp\left(-\frac{(f-g_r)^2}{2\sigma_{l,r}^2}\right) df \text{ for } f < g_r.$$

(1)

By finding the value of f at which the product of the pdf's is a maximum, we may determine the maximum-likelihood value of f, which we denote by $g_{ML}$.

Given $g_{ML}$ and the error estimates, we may form pdf's for the expected values of the measurements $g_r$, on the assumption that $f = g_{ML}$:

$$p_r(g_r)dg_r = \frac{2^{1/2}}{\pi^{1/2}(\sigma_{u,r}+\sigma_{l,r})} \exp\left(-\frac{(g_r-g_{ML})^2}{2\sigma_{l,r}^2}\right) dg_r \text{ for } g_r > g_{ML},$$

$$p_r(g_r)dg_r = \frac{2^{1/2}}{\pi^{1/2}(\sigma_{u,r}+\sigma_{l,r})} \exp\left(-\frac{(g_r-g_{ML})^2}{2\sigma_{u,r}^2}\right) dg_r \text{ for } g_r < g_{ML}.$$

(2)

(Note the reversal of the roles of the two error estimates.)

From this pdf, we may form the cumulative probability distribution function (cdf):

$$C_r(g_r) = \int_{-\infty}^{g_r} dx\, p_r(x). \tag{3}$$

If we go through a similar procedure for a variable $y$ and measurements $Y_r$, except that the pdf is now normal, centered on zero with width unity, then to any value $Y$ there corresponds a cdf value given by

$$H = \int_{-\infty}^{Y} dy \frac{1}{(2\pi)^{1/2}} \exp\left(-\tfrac{1}{2}y^2\right). \tag{4}$$

The important point is that, based on the assumption of Equation (1), the cdf values $C_r$ are expected to be uniformly distributed over the range 0 to 1. Similarly, on the assumption that *y* has a normal distribution, the cdf value *H* also is uniformly distributed over the same range. Hence we may now "map" each value $g_r$ onto a "proxy" variable $y_r$, by finding the value of *y* for which $H = C_r$. We may regard this proxy variable as a normalized form of the variable $g_r$,

## 4. Power Spectrum Analysis

Since there appear to be systematic differences between the four experiments in the GALLEX series, we carry out the above procedure for each experiment, and then concatenate the results to form a complete normalized time-series $y_r$ for all 65 runs of the GALLEX series. Before proceeding to time-frequency analysis of the normalized variable, we show the result of power spectrum analyses of the normalized variables for GALLEX and for GNO.

The power spectrum derived from the normalized GALLEX data is shown in Figure 2. In Table 2, we list the top ten peaks in the range $0 - 30$ yr$^{-1}$. The power spectrum derived from the normalized GNO data is shown in Figure 3, and the top ten peaks in the range $0 - 30$ yr$^{-1}$ are listed in Table 3.

We see from Figures 2 and 3 that the power spectra derived from the normalized data are similar to those previously obtained (Sturrock, Caldwell, and Scargle 2006). In the GALLEX power spectrum, the peaks are somewhat higher than in the earlier version. The GNO power spectrum shows comparatively little evidence of modulation. For this reason, we focus primarily on GALLEX data in the remainder of this article.

In the GALLEX power spectrum, the strongest peak is that at 13.63 yr$^{-1}$ with power 7.98, and the second strongest is that at 11.87 yr$^{-1}$ with power 5.47. As we noted in the Introduction, these frequencies sum to 25.50 yr$^{-1}$, corresponding to a period of 14.32 days, close to two weeks, which raises the question of whether one is an alias of the other. In order to pursue this question, we shall return to time-frequency analysis in Section 6.

## 5. Power Spectrum of GALLEX Timing Data

In order to understand aliasing in the power spectrum, it is helpful to form a power spectrum of the timing data. We may construct this by computing the Rayleigh power of a time series formed from the mean times of the runs of the GALLEX series. This is given simply by

$$S_T = \frac{1}{R}\left|\sum_{r=1}^{R}\exp(i2\pi\nu t_{m,r})\right|^2, \qquad (5)$$

where R is the number of runs, and $t_{m,r}$ is the mean time of each run. Since we shall be examining power spectra up to 30 $yr^{-1}$, we need to compute the power spectrum of the timing data up to 60 $yr^{-1}$. The result is shown in Figure 4, and the top ten peaks are listed in Table 4.

## 6. Time-Frequency Analysis of the Normalized Data

We now repeat the time-frequency analysis of Section 2, applying the procedure to the normalized GALLEX data derived in Section 3. The result is shown in Figure 5. We see that the display is similar to that of Figure 1, but the sharp discontinuity at about 1994 has nearly disappeared. It appears that we have effectively merged the data for the four GALLEX experiments.

We also show in Figure 6, for comparison, the result of a time-frequency analysis of the GNO dataset. We see that there are actually strong features in this figure also. Viewed from this perspective, the difference between the GALLEX and GNO datasets is not so much that GALLEX shows evidence of modulation but GNO does not; it is more that modulation in the GALLEX dataset tends to be stable in frequency, whereas that in the GNO dataset tends to drift in frequency.

We may discern in both Figure 5 and Figure 6 a clear tendency for the displays to be mirror symmetric with respect to the frequencies 26 yr$^{-1}$, 13 yr$^{-1}$, and 6.5 yr$^{-1}$. These correspond approximately to periods of two weeks, four weeks, and 8 weeks. A pair of features that are symmetric with respect to 26 yr$^{-1}$ have frequencies that sum to 52 yr$^{-1}$, corresponding to a period of one week, etc. Due to the fact that GALLEX and GNO data were compiled in runs that are a few (typically 3 or 4) weeks in duration (and in view of Figure 4), it is not surprising that we should find such aliasing.

Modulations at about 11.8 yr$^{-1}$ and 13.6 yr$^{-1}$ are clearly visible in Figure 5, beginning in about 1993.7 and continuing until about 1995.7. The facts that they are similar in strength and in duration, and that they are more or less symmetric with respect to 13 yr$^{-1}$, strongly suggest that they are an alias pair. This poses the question – which is the true modulation, and which is the alias? We take up this question in the next section.

In Figure 6 (for GNO data), we see modulation at several frequencies. However, the most interesting is that near 9.43 yr$^{-1}$, which is the frequency of the principal peak that we have found (Sturrock 2004; Sturrock and Scargle 2006) in the Super-Kamiokande dataset (Fukuda et al., 2001, 2002, 2003). We see feature also near 4 yr$^{-1}$, 17 yr$^{-1}$, 22 yr$^{-1}$, and 30 yr$^{-1}$. These could all be aliases of the peak that drifts in frequency over the range 9 to 10 yr$^{-1}$.

## 7. Time-Frequency Analysis of Principal Modulations

Returning to GALLEX, we have determined the complex amplitudes for the frequencies 11.87 yr$^{-1}$ and 13.63 yr$^{-1}$, which are the two principal peaks in the power spectrum, derived in Section 4, of the normalized time series $y_r$ that we derived in Section 3. We have then formed time-frequency displays, replacing the actual normalized variable with the sinusoidal variables corresponding to the two complex amplitudes. These displays are shown in Figures 7 and 8, respectively.

On comparing these two figures with Figure 5, the time-frequency display for the normalized data, it is obvious that modulation at either 11.87 yr$^{-1}$ or at 13.63 yr$^{-1}$ leads to features at both frequencies, and also to features near 1 yr$^{-1}$, 25 yr$^{-1}$, and 27 yr$^{-1}$. Features near 25 yr$^{-1}$, and 27 yr$^{-1}$ are obvious in Figure 5, but there is only a hint of a feature near 1 yr$^{-1}$.

It is worth noticing that the features near 6 yr$^{-1}$ and 7 yr$^{-1}$ in Figure 5 are not replicated in either Figure 7 or Figure 8, indicating that they are not due to aliasing of modulation at either 11.87 yr$^{-1}$ or 13.63 yr$^{-1}$. Since these frequencies, listed as 6.05 yr$^{-1}$ and 6.93 yr$^{-1}$ in Table 2, sum to 12.98 yr$^{-1}$ (corresponding to a period of 28.14 days or about 2 weeks), it is clear that these comprise an alias pair. We comment further on this pair in Section 8.

We have also carried out time-frequency analysis of the normalized data, *subtracting* the sinusoidal variables corresponding to the complex amplitudes for the two frequencies 11.87 yr$^{-1}$ and 13.63 yr$^{-1}$. The resulting displays are shown in Figures 9 and 10, respectively. We see that subtracting out the 11.87 yr$^{-1}$ modulation removes almost all features at or near 11.87 yr$^{-1}$ and 13.63 yr$^{-1}$, and also removes the structure in the range 24 - 28 yr$^{-1}$. On the other hand, subtracting out the 13.63 yr$^{-1}$ modulation removes much of the structure at or near 11.87 yr$^{-1}$ and 13.63 yr$^{-1}$, but leaves visible structure at those frequencies and also in the range 24 - 28 yr$^{-1}$ after 1995.0.

The comparison is not conclusive but the evidence, such as it is, points towards 11.87 yr$^{-1}$ as the primary modulation, of which the modulation at 13.63 yr$^{-1}$ is an alias.

We have repeated these two analyses for GNO data. This is not so promising, since the peaks in Figure 6 drift in frequency. As a compromise, we adopt the value 9.20 yr$^{-1}$ listed in Table 3. We select this peak since it is close to the frequency of the principal modulation (at 9.43 yr$^{-1}$) in the Super-Kamiokande dataset (Sturrock 2004; Sturrock and Scargle 2006). The time-frequency display formed by replacing the actual normalized variable with the sinusoidal variable corresponding to the complex amplitude at 9.20 yr$^{-1}$ is shown in Figure 11. The display formed by subtracting this modulation from the actual (normalized) GNO dataset is shown in Figure 12. We see that the pattern in Figure 11 is similar to that in Figure 6. Furthermore, we from Figure 12 that subtracting the modulation at 9.20 yr$^{-1}$ removes most of the features after 2001.0. The fact that features remain for times less than 2001.0 may be attributed to the fact that the modulation drifts in frequency.

## 8. Discussion

In our earlier articles, (Sturrock, Scargle, Walther, and Wheatland, 1999; Sturrock and Scargle, 2001; Sturrock and Weber, 2002; Sturrock, Caldwell, and Scargle, 2006), we considered the modulation at or near 13.63 yr$^{-1}$ to be the most prominent. Since this is the synodic rotation frequency in an equatorial section of the Sun at radius close to 0.8 $R_{solar}$ (Schou et al. 2002), it seemed reasonable to attribute this periodicity to modulation of the solar neutrino flux occurring in the convection zone. This led us to consider the possibility that electron neutrinos might be converted to mu or tau neutrinos, or to a sterile neutrino, by RSFP (Resonant Spin-Flavor Process; Akhmedov 1988; Lim and Marciano 1998; Caldwell 2007). This requires that neutrinos have a transition magnetic dipole moment. Calculations by Chauhan, Pulido, and Raghavan, 2005) have shown that this model can explain the apparent rotational modulation in GALLEX data, and also improve the theoretical fit of time-averaged data from all experiments.

On the other hand, there is a puzzle if we pursue the possibility that the true modulation is that at 11.87 yr$^{-1}$. If this were the "synodic" rotation frequency where modulation occurs, this would imply that the sidereal frequency is 12.87 yr$^{-1}$. We may compare this estimate with estimates of the internal rotation rate of an equatorial section of the Sun derived from helioseismology, shown in Figure 13, which combines estimates from MDI data (Schou et al. 2002) and more recent estimates from GONG data (Garcia et al. 2007). We see that 12.87 yr$^{-1}$ is too low a frequency to be compatible either with the convection zone or with most of the radiative zone. For r < 0.2 $R_{solar}$, the rotation rate is not well determined by helioseismological observations, so modulation at this frequency could conceivably be occurring deep in the core.

However, the neutrino-flux modulation at 11.87 yr$^{-1}$ need not imply that modulation is occurring where the sidereal rotation rate is 12.87 yr$^{-1}$. We have shown (Sturrock and Bai 1993) that other sidebands (other than the first lower sideband) can show up in a power spectrum of measurements derived from a rotating model of the Sun that has a complex structure and an axis that is not normal to the ecliptic. We may therefore consider the possibility that the peak at 11.87 yr$^{-1}$ corresponds to the *second* lower sideband of the rotation frequency, which then becomes 13.87 yr$^{-1}$. We see from Figure 13 that this value is consistent with the internal rotation rate in the radiative zone just below the tachocline (which we take to be located at 0.7 R$_{solar}$) or in the core, but not in the convection zone.

We find supporting evidence for this choice of the sidereal rotation frequency in our analysis of Homestake data (Sturrock, Walther and Wheatland, 1997), where we find a prominent peak in the power spectrum at 12.88 yr$^{-1}$, which we have interpreted as the synodic frequency (the first lower sideband) corresponding to a sidereal rotation frequency of 13.88 yr$^{-1}$. We see that, in this respect, there is virtually perfect agreement between our present analysis of GALLEX data and our earlier analysis of Homestake data.

We found additional periodicities in the Homestake data at 10.83 yr$^{-1}$, 11.85 yr$^{-1}$, 13.85 yr$^{-1}$, and 14.88 yr$^{-1}$., which may be interpreted as two lower sidebands, the fundamental, and one upper sideband. We also drew attention to a peak at frequency 2.32 yr$^{-1}$, corresponding to a period of 157 days. This is close to the period of the "Rieger" oscillation (period 153 days, frequency 2.40 yr$^{-1}$) that was first discovered in the timing of gamma-ray flares (Rieger et al. 1984), but occurs also in other indices of solar activity. Similar modulations of solar activity have since been discovered at other frequencies. Bai (2003) has drawn attention to the periods 76 days and 51 days, corresponding to frequencies 4.80 yr$^{-1}$ and 7.16 yr$^{-1}$. We find that these frequencies are all compatible with our proposed interpretation in terms of r-modes (Sturrock Scargle, Walther, and Wheatland 1999).

In a fluid sphere that rotates with frequency $v_R$, r-modes (Papaloizou and Pringle, 1978; Provost, Berthomieu, and Rocca, 1981; Saio 1982) are retrograde (with respect to the rotating fluid) waves with frequencies

$$\nu(l,m \mid Sun) = \frac{2m\nu_R}{l(l+1)}, \qquad (6)$$

where l and m are two of the three spherical-harmonic indices (the frequency is independent of n). The index l takes the values $l = 2, 3,...$, and m takes the values $m = 1,...,l$. As seen from inertial space, their frequencies are

$$\nu(l,m \mid inertial) = m\nu_R - \frac{2m\nu_R}{l(l+1)}, \qquad (7)$$

and as seen from Earth, and for waves rotating in the plane of the ecliptic, the frequencies (measured in cycles per year) are

$$\nu(l,m \mid Earth) = m(\nu_R - 1) - \frac{2m\nu_R}{l(l+1)}. \qquad (8)$$

The fact that the frequency is independent of n implies that r-mode oscillations may have any radial profile: in particular, the oscillation may be confined to a narrow range of radius.

The above three Rieger-type periodicities are consistent with estimates given by Equation (6) for the choice l = 3, m = 1, 2, and 3, if we adopt a sidereal rotation rate 14.30 $yr^{-1}$. We see from Figure 13 that this corresponds to the sidereal rotation rate in the vicinity of (slightly above) the tachocline. The fact that frequencies given by Equation (6) show up in indices of solar activity implies that r-modes are interacting with one or more structures (possibly magnetic) that have fixed locations in the rotating Sun.

We find similar, but not identical, frequencies in power spectra formed from solar neutrino data. The periodicity at 2.32 $yr^{-1}$, found in Homestake data, may be interpreted as an r-mode oscillation with l = 3, m = 3, for a sidereal rotation rate of 13.92 $yr^{-1}$, close to the value we have found in our current analysis. If we adopt 13.87 $yr^{-1}$ as the sidereal rotation frequency, then Equation (6) yields 2.33 yr$^{-1}$, 4.62 yr$^{-1}$ and 6.94 yr$^{-1}$ for the frequencies of the l = 3, m = 1, 2 and 3, r-modes. We see from Table 2 that the GALLEX power spectrum shows peaks at 4.54 yr$^{-1}$ and at 6.93 yr$^{-1}$. In discussing GNO data, we drew attention to the peak at 9.20 yr-1, since this is close to the principal peak (at 9.43 yr$^{-1}$) in the Super-Kamiokande power spectrum, which we proposed may be interpreted as the l = 2, m = 2 r-mode frequency (Sturrock 2004). For a sidereal rotation rate of 13.87 yr$^{-1}$. Equation (6) yields the value 9.24 yr$^{-1}$ for the frequency of this mode. In Table 5, we show a list of these frequencies and the frequencies to be expected on the basis of Equation (6) if the r-modes occur where the sidereal rotation frequency is 13.87 yr$^{-1}$.

We saw from Figure 13 that we may relate the results of Sections 6 and 7 to solar processes if we attribute those results to modulation in a region of the Sun where the sidereal rotation rate is 13.87 yr$^{-1}$. We now see that the additional periodicities noted in Table 5 may also be attributed to processes where the sidereal rotation rate has this value. This estimate is compatible with the rotation rate in the outer region of the convection zone, but it is also consistent with current knowledge about the rotation rate of the core.

It appears, therefore, that our analysis leaves us with two options: One is that the variability of the solar neutrino flux has its origin in the outer region of the radiative zone (say in the range 0.6 to 0.7 R$_{solar}$). The other option is that the variability has its origin in the core, which would imply that nuclear burning is not spherically symmetric and is not steady.

In either case, we will need to understand the process responsible for the excitation of r-modes. It is possible that the excitation of r-modes just above the tachocline (those responsible for the Rieger oscillations) and just below the tachocline (those responsible for the periodicities noted in Table 5) are due to the velocity gradients in those regions. For the latter option, it may be that the excitation of r-modes is due to a symbiotic relationship between these oscillations and conditions determining the rate of nuclear burning.

However, the former process would require that neutrinos have a transition magnetic moment and that a dynamo-generated magnetic field extends below the tachocline as well as above it. Current dynamo theory (for a recent review, see Dikpati 2005) implies that the radiative zone is substantially free from magnetic field. If this is the case, or if it is established that neutrinos have no significance transition magnetic moment, we shall be led to attribute variability of the solar neutrino flux to processes in the core. If this is the case, we may infer from our analysis that the core is in substantially rigid rotation with a sidereal rotation frequency close to 13.87 $yr^{-1}$, i.e. 440 nHz.

## Acknowledgements

We wish to thank Rachel Howe, Alexander Kosovichev, John Leibacher, and Joao Pulido, for helpful discussions related to this work, which was supported by NSF Grant AST-0607572.

**Figure Titles**

Figure 1. Time-Frequency analysis for GALLEX, obtained by applying a likelihood procedure to sliding sections of 20 runs each.

Figure 2. Power spectrum analysis of normalized GALLEX measurements.

Figure 3. Power spectrum analysis of normalized GNO measurements.

Figure 4. Power spectrum formed from the mean times of the GALLEX runs.

Figure 5. Time-Frequency analysis for GALLEX, obtained by applying a likelihood procedure to the normalized measurements derived in Section 3.

Figure 6. Time-Frequency analysis for GNO, obtained by applying a likelihood procedure to the normalized measurements derived in Section 3.

Figure 7. Time-frequency analysis for GALLEX time sequence, replacing the normalized measurements with the best-fit sinusoidal modulation at 11.87 $yr^{-1}$.

Figure 8. Time-frequency analysis for GALLEX time sequence, replacing the normalized measurements with the best-fit sinusoidal modulation at 13.63 $yr^{-1}$.

Figure 9. Time-frequency analysis for GALLEX time sequence, after subtracting the best-fit sinusoidal modulation at 11.87 $yr^{-1}$ from the normalized measurements.

Figure 10. Time-frequency analysis for GALLEX time sequence, after subtracting the best-fit sinusoidal modulation at 13.63 $yr^{-1}$ from the normalized measurements.

Figure 11. Time-frequency analysis for GNO time sequence, replacing the normalilzed variable with the best-fit sinusoidal modulation at 9.20 $yr^{-1}$.

Figure 12. Time-frequency analysis for GNO time sequence, after subtracting the best-fit sinusoidal modulation at 9.20 $yr^{-1}$ from the normalized variable.

Figure 13. Solar internal rotation rates estimated on the basis of MDI data (green) and GONG data (blue), possible rotation rates inferred from GALLEX-GNO data (red), and rotation rate inferred from Rieger-type oscillations (magenta).

TABLE 1

GALLEX Data

| Runs | Mean Count Rate | Standard Deviation |
|---|---|---|
| 1 - 15 | 85.3 | 75.5 |
| 16 - 39 | 75.1 | 37.2 |
| 40 - 53 | 49.5 | 58.6 |
| 54 - 65 | 108.2 | 71.2 |

TABLE 2

Top Ten Peaks in the GALLEX Power Spectrum

| Frequency ($yr^{-1}$) | Power |
|---|---|
| 13.63 | 7.98 |
| 11.87 | 5.47 |
| 4.54 | 5.46 |
| 13.07 | 5.4 |
| 6.05 | 5.21 |
| 27.32 | 5.14 |
| 6.93 | 4.83 |
| 3.97 | 4.61 |
| 21.55 | 4.31 |
| 12.26 | 4.11 |

TABLE 3

Top Ten Peaks in the GNO Power Spectrum

| Frequency (yr$^{-1}$) | Power |
|---|---|
| 21.93 | 5.45 |
| 2.93 | 4.80 |
| 3.52 | 4.25 |
| 9.20 | 4.10 |
| 17.29 | 4.00 |
| 16.53 | 3.70 |
| 15.92 | 3.54 |
| 10.07 | 3.30 |
| 1.10 | 3.18 |
| 23.03 | 2.98 |

TABLE 4

Top Ten Peaks in the Power Spectrum of End Times of GALLEX Runs

| Frequency (yr$^{-1}$) | Power |
|---|---|
| 12.99 | 18.12 |
| 51.98 | 17.70 |
| 39.22 | 17.65 |
| 26.26 | 12.30 |
| 52.36 | 10.92 |
| 25.90 | 10.10 |
| 17.01 | 8.28 |
| 16.60 | 7.70 |
| 33.62 | 7.49 |
| 16.81 | 7.23 |

TABLE 5

Periodicities in the solar-neutrino flux interpreted as r-mode oscillations.

| Experiment | l | m | Found Frequency | Expected Frequency |
|---|---|---|---|---|
| Homestake | 3 | 1 | 2.32 yr$^{-1}$ | 2.31 yr$^{-1}$ |
| GALLEX | 3 | 2 | 4.54 yr$^{-1}$ | 4.62 yr$^{-1}$ |
| GALLEX | 3 | 3 | 6.93 yr$^{-1}$ | 6.93 yr$^{-1}$ |
| GNO | 2 | 2 | 9.20 yr$^{-1}$ | 9.25 yr$^{-1}$ |

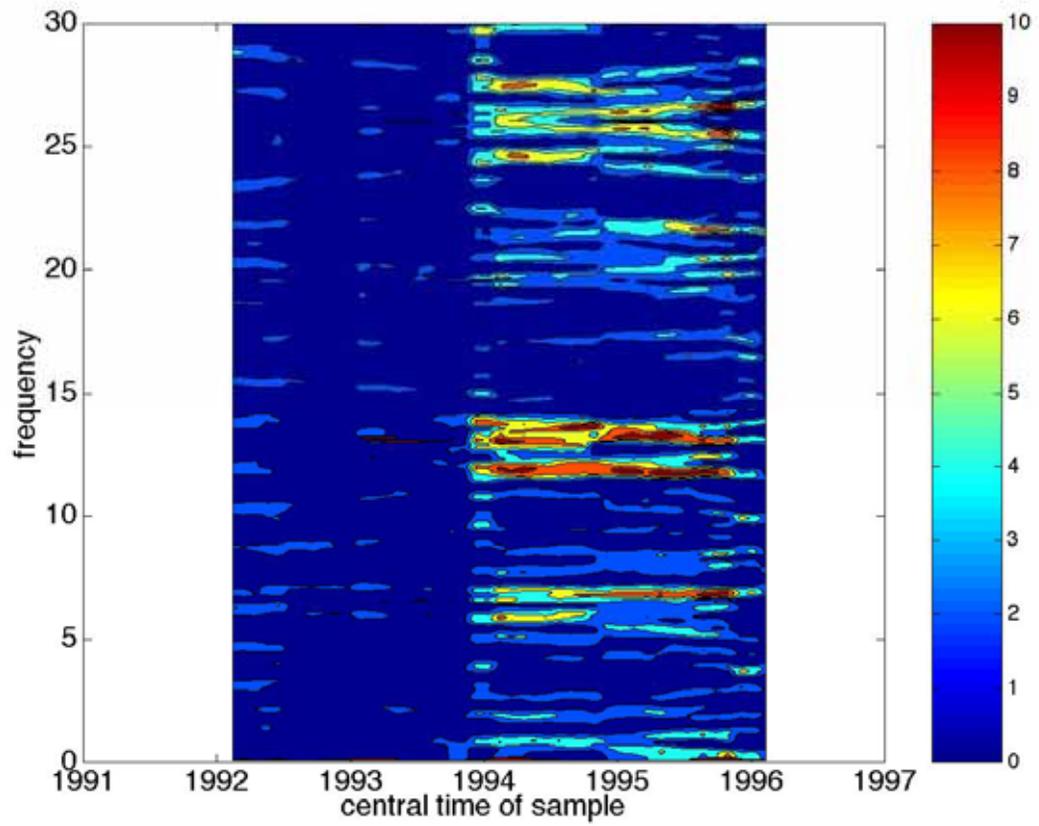

Figure 1. Time-Frequency analysis for GALLEX, obtained by applying a likelihood procedure to sliding sections of 20 runs each.

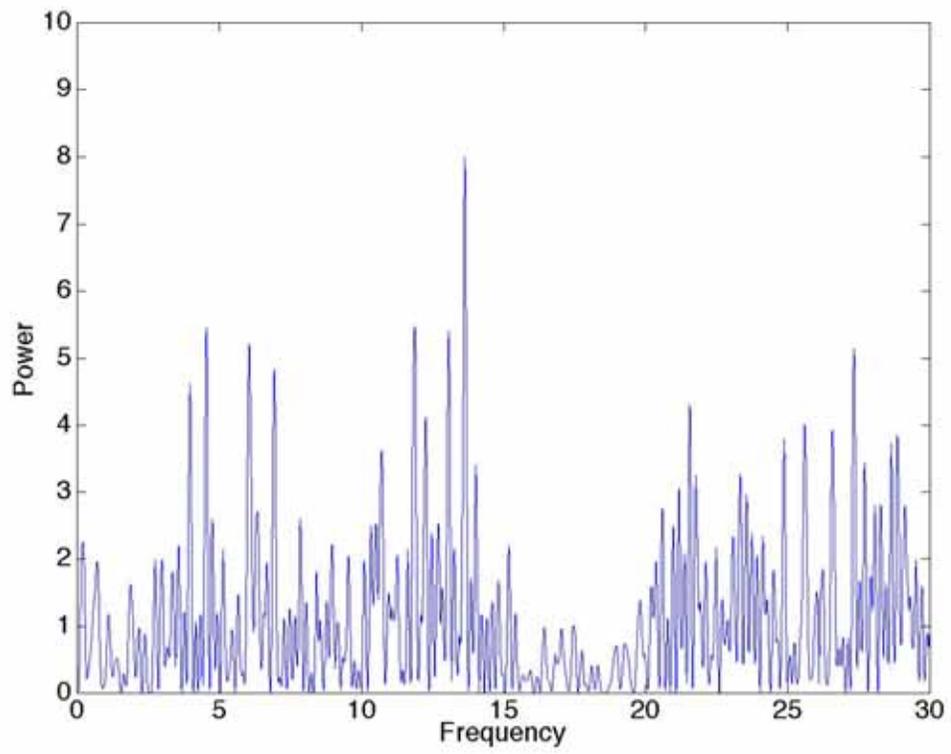

Figure 2. Power spectrum analysis of normalized GALLEX measurements.

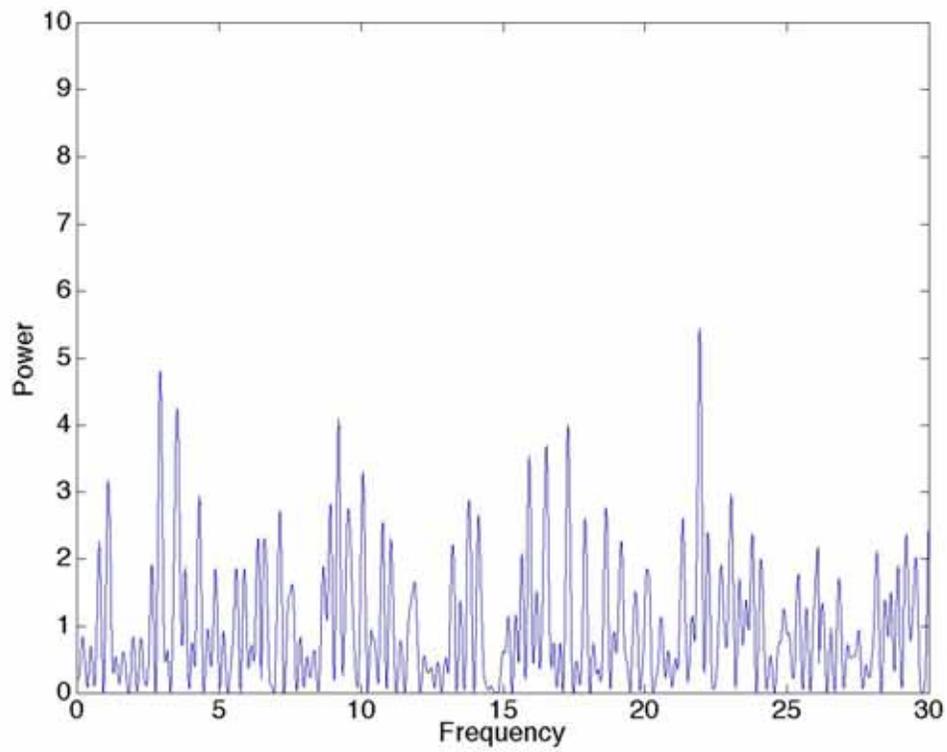

Figure 3. Power spectrum analysis of normalized GNO measurements.

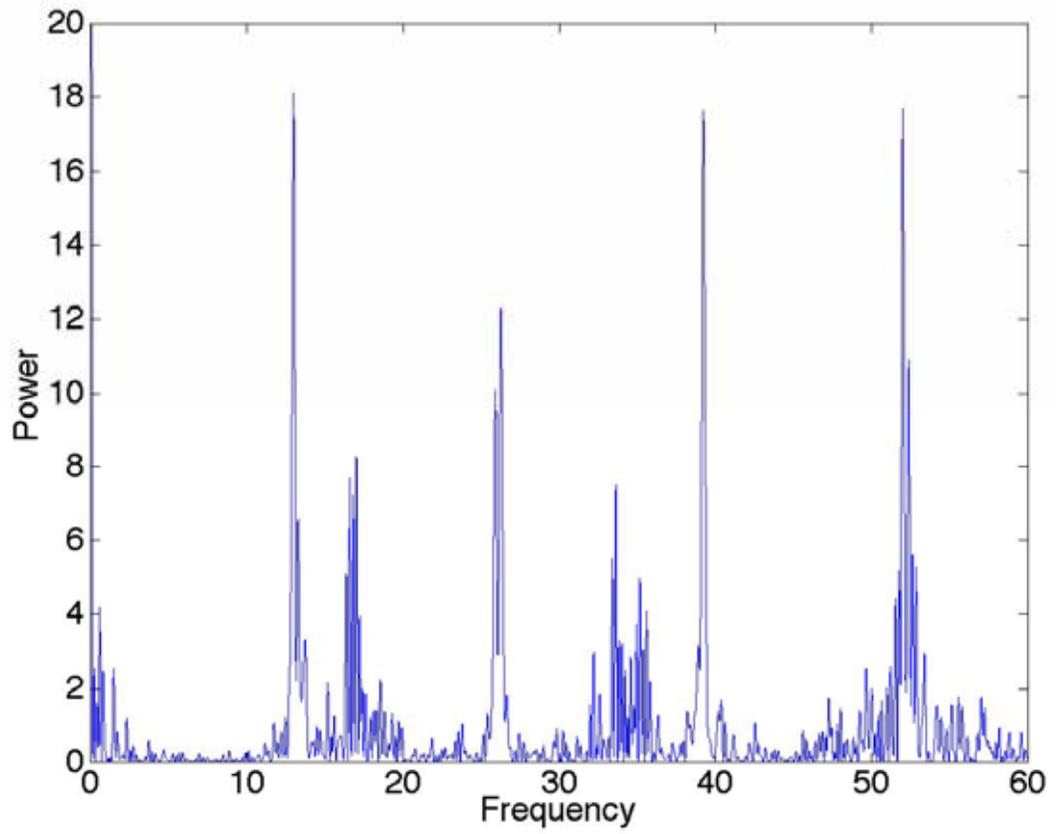

Figure 4. Power spectrum formed from the mean times of the GALLEX runs.

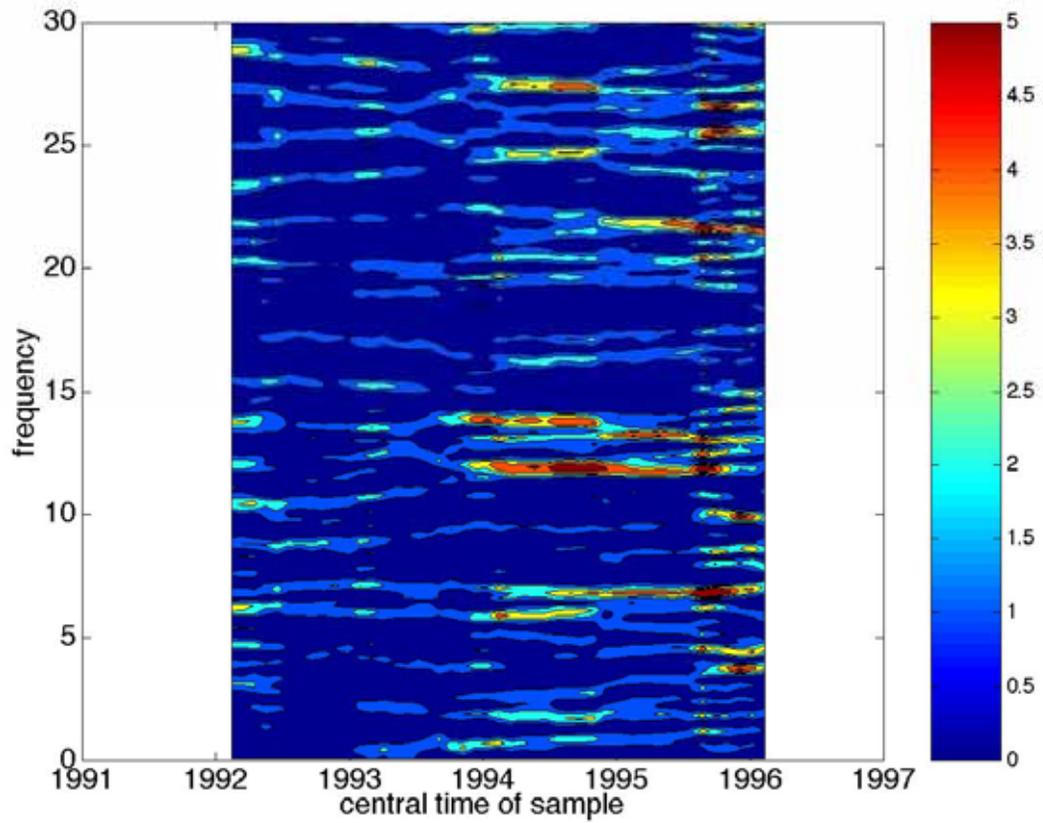

Figure 5. Time-Frequency analysis for GALLEX, obtained by applying a likelihood procedure to the normalized measurements derived in Section 3.

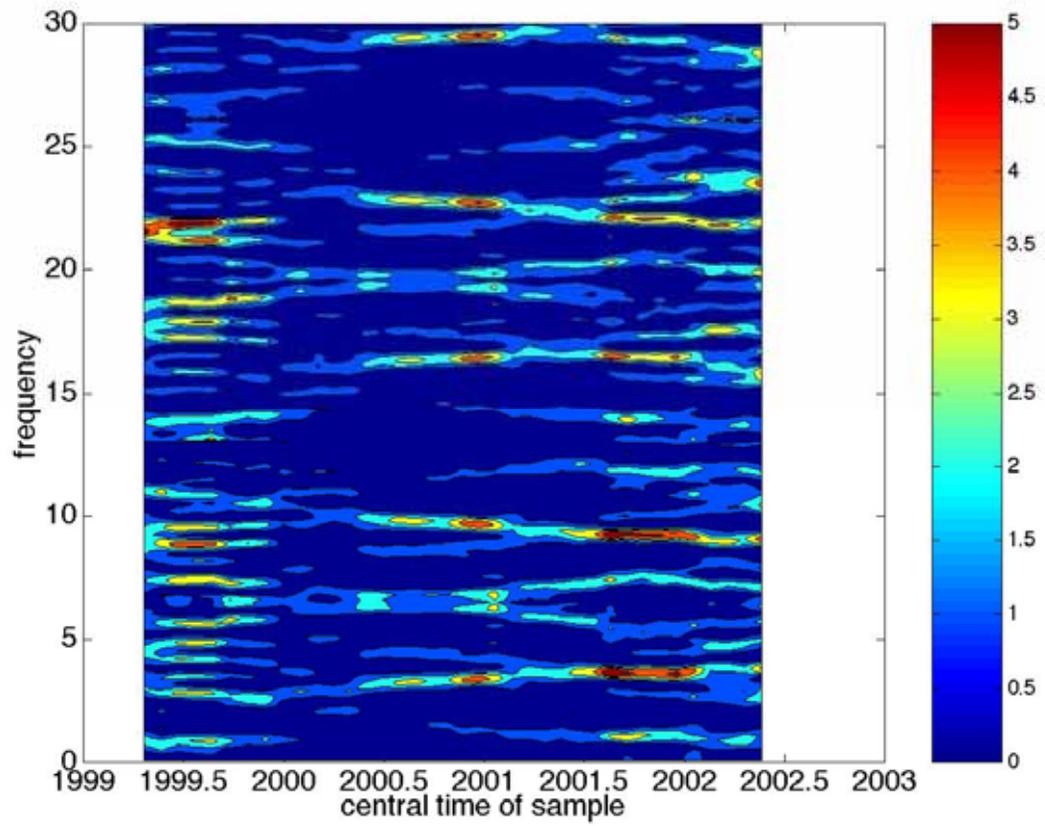

Figure 6. Time-Frequency analysis for GNO, obtained by applying a likelihood procedure to the normalized measurements derived in Section 3.

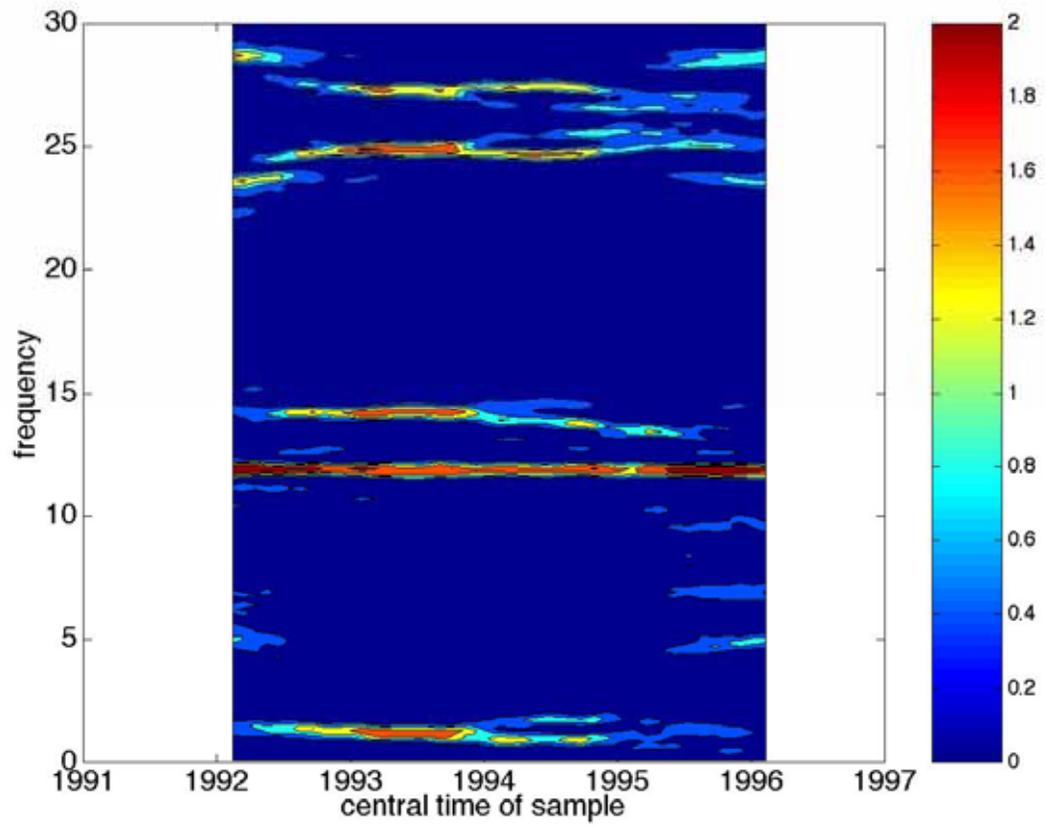

Figure 7. Time-frequency analysis for GALLEX time sequence, replacing the normalized measurements with the best-fit sinusoidal modulation at 11.87 yr$^{-1}$.

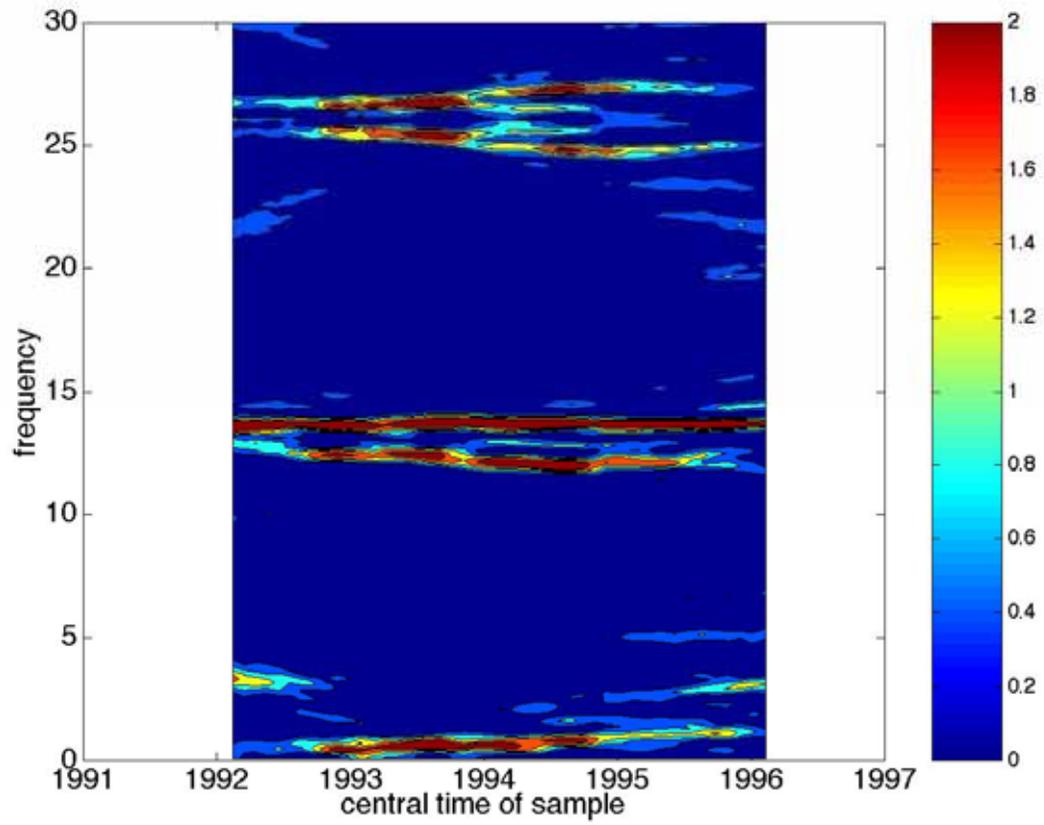

Figure 8. Time-frequency analysis for GALLEX time sequence, replacing the normalized measurements with the best-fit sinusoidal modulation at 13.63 yr$^{-1}$.

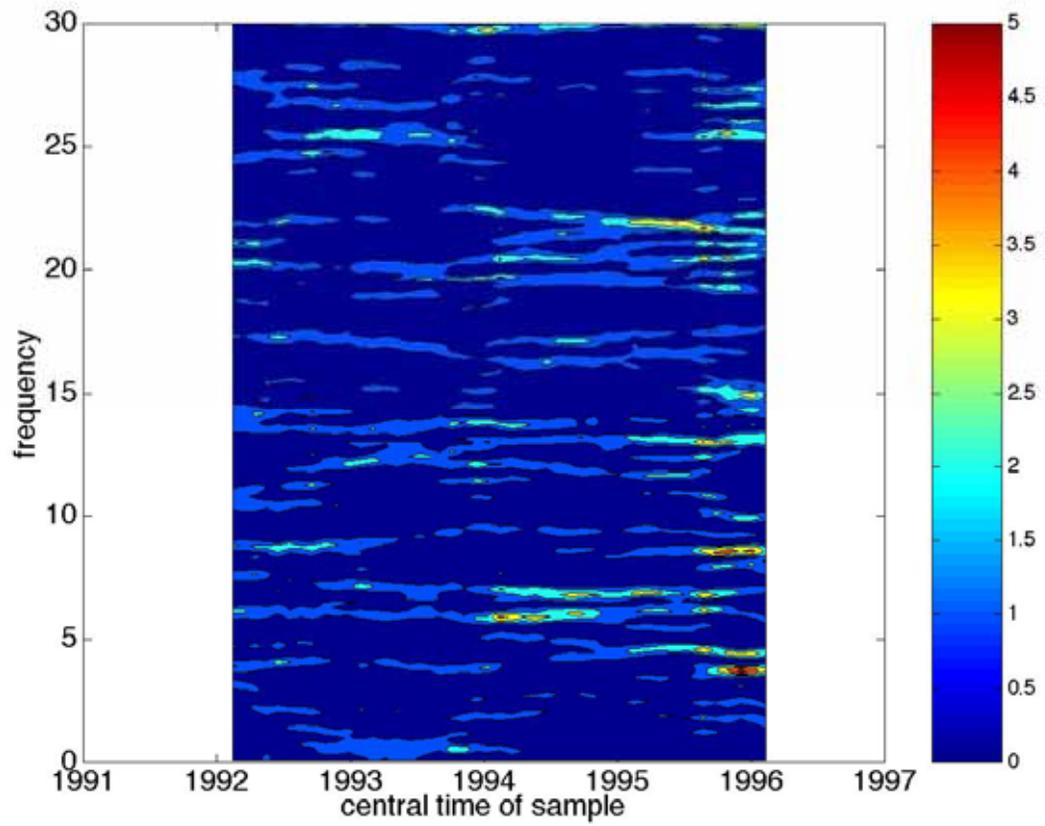

Figure 9. Time-frequency analysis for GALLEX time sequence, after subtracting the best-fit sinusoidal modulation at 11.87 yr$^{-1}$ from the normalized measurements.

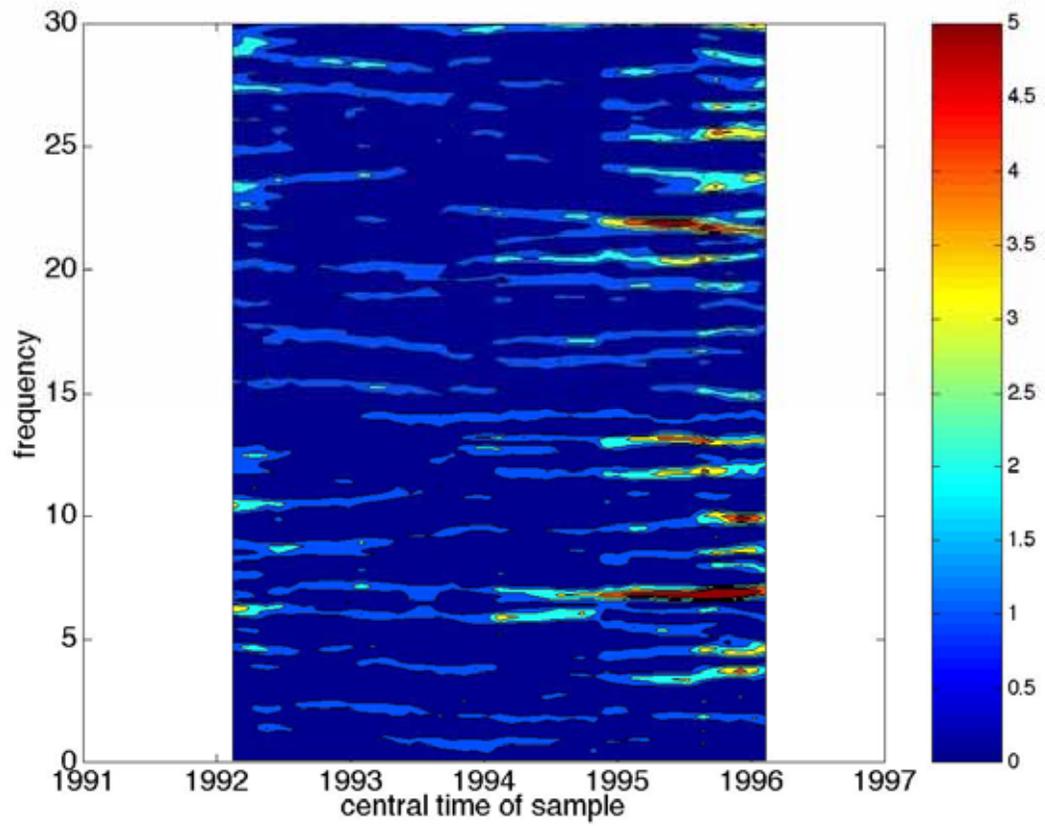

Figure 10. Time-frequency analysis for GALLEX time sequence, after subtracting the best-fit sinusoidal modulation at 13.63 yr$^{-1}$ from the normalized measurements.

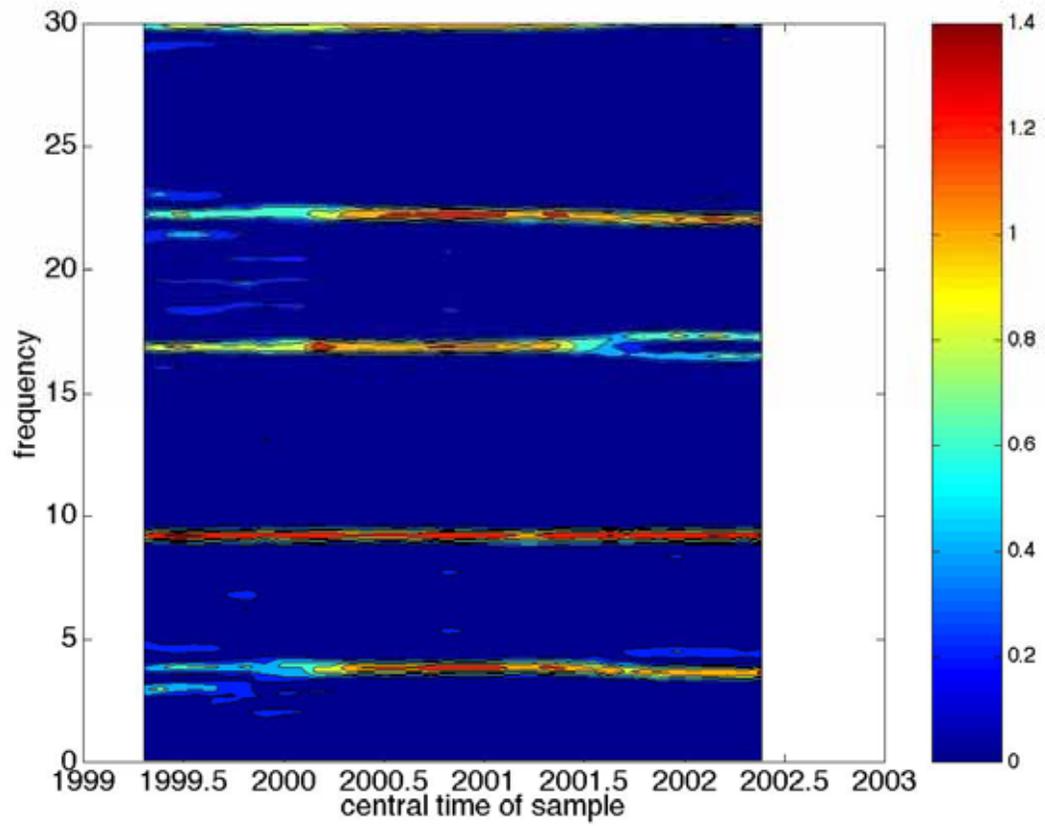

Figure 11. Time-frequency analysis for GNO time sequence, replacing the normalized variable with the best-fit sinusoidal modulation at 9.20 yr$^{-1}$.

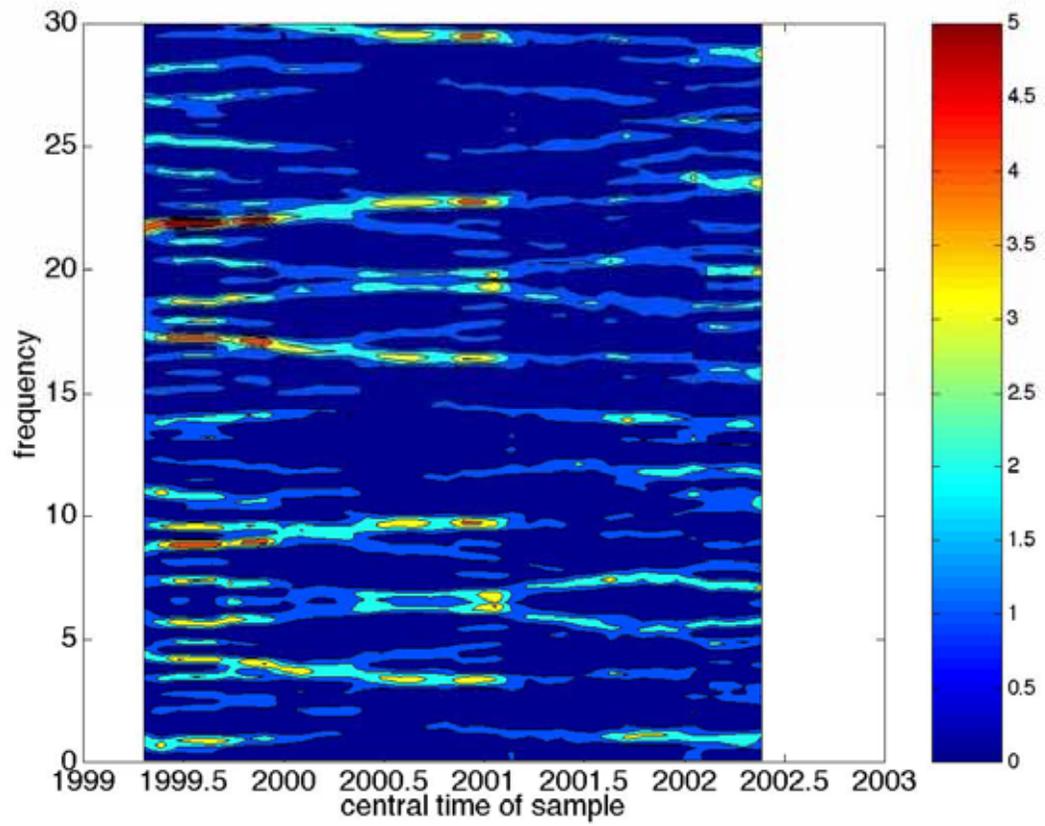

Figure 12. Time-frequency analysis for GNO time sequence, after subtracting the best-fit sinusoidal modulation at 9.20 yr$^{-1}$ from the normalized variable.

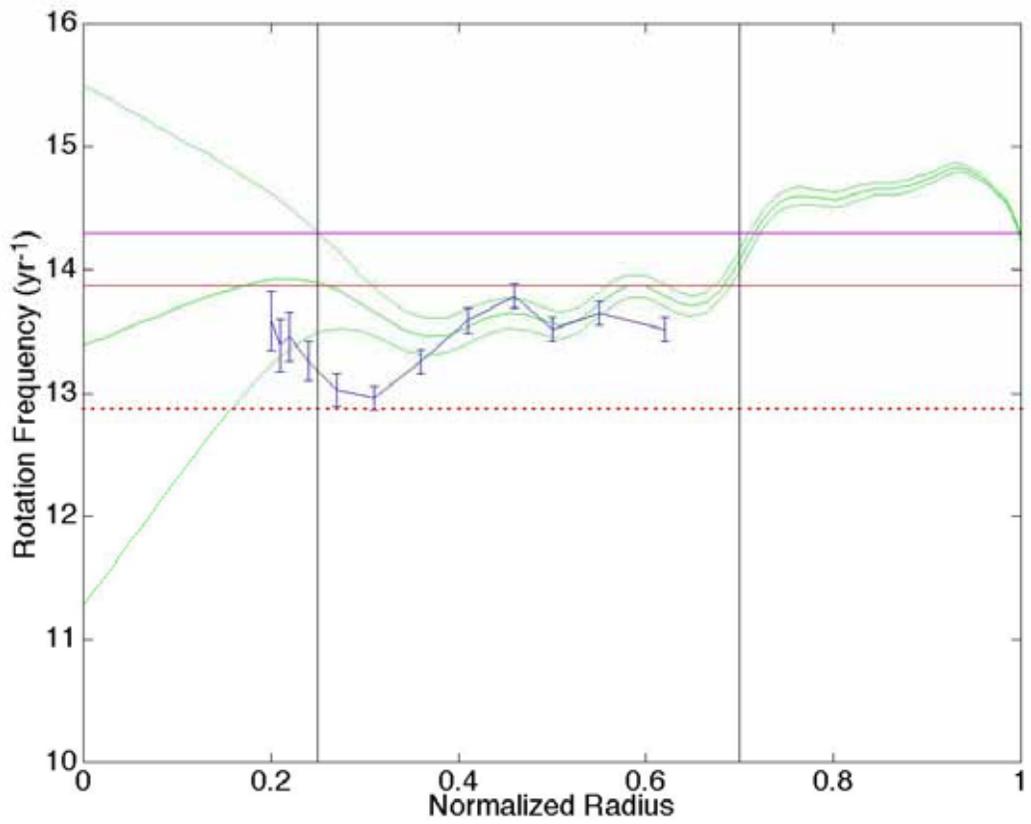

Figure 13. Solar internal rotation rates estimated on the basis of MDI data (green) and GONG data (blue), possible rotation rates inferred from GALLEX-GNO data (red), and rotation rate inferred from Rieger-type oscillations (magenta).